\documentclass{PoS}

\usepackage{amsmath}
\usepackage{gensymb}

\title{Chiral symmetry breaking corrections to the pseudoscalar pole contribution of the Hadronic Light-by-Light piece of $a_\mu$}

\ShortTitle{Chiral symmetry breaking corrections to $a_\mu^{P,HLbL}$}

\author{\speaker{Adolfo Guevara}$^{a}$, Pablo Roig$^b$ and Juan Jos\'e Sanz Cillero$^a$ 
\\
        \llap{$^a$} Departamento de F\'isica Te\'orica and UPARCOS, Universidad Complutense de Madrid
        Plaza de Ciencias 1, Facultad de CC F\'isicas, 28040 Madrid, Spain\\
        \llap{$^b$} Departamento de F\'isica, Centro de Investigaci\'on y de Estudios Avanzados del IPN, 
        Apartado Postal 14-740, 07000, Mexico City, Mexico\\
        E-mail: \email{adguevar@ucm.es}, \email{jjsanzcillero@ucm.es}, \email{proig@fis.cinvestav.mx}}

\abstract{We have studied the $P\to\gamma^\star\gamma^\star$ form factor in Resonance Chiral Theory, with 
  $P = \pi^0\eta\eta'$, to 
  compute the contribution of the pseudoscalar pole to the hadronic light-by-light piece of the 
  anomalous magnetic moment of the muon. In this work we allow the leading $U(3)$ chiral symmetry
  breaking terms, obtaining the most general expression for the form factor up to $\mathcal{O}(m_P^2)$.
  The parameters of the Effective Field Theory are obtained by means of short distance constraints 
  on the form factor and matching with the expected behavior from QCD. Those parameters that cannot 
  be fixed in this way are fitted to experimental determinations of the form factor within the spacelike 
  region. Chiral symmetry relations among the transition form factors 
  for $\pi^0,\eta$ and $\eta'$ allow for a simultaneous fit to experimental data for the three mesons. 
  This shows an inconsistency between the BaBar $\pi^0$ data and the rest of the experimental inputs. 
  Thus, we find a total pseudoscalar pole contribution of $a_\mu^{P,HLbL}=(8.47\pm 0.16)\cdot 10^{-10}$
  for our best fit (that neglecting the BaBar $\pi^0$ data). Also, a preliminary rough estimate of the 
  impact of NLO in $1/N_C$ corrections and higher vector multiplets (asym) enlarges the uncertainty up to
  $a_\mu^{P,HLbL}=(8.47\pm 0.16_{\rm stat}\pm 0.09_{N_C}{}^{+0.5}_{-0.0_{\rm asym}})10^{-10}$. This contribution is 
  based on our work in ref. \cite{Guevara:2018rhj}.}

\FullConference{XIII Quark Confinement and the Hadron Spectrum - Confinement2018\\
		31 July - 6 August 2018\\
		Maynooth University, Ireland}
\begin{document}

\section{Introduction}

\label{intro}
  The intrinsic magnetic moment of particles is an outstanding observable, thanks to 
  the first measurement of the magnetic moment of silver atoms in Stern-Gerlach experiments, the 
  non-commutative nature of angular momentum was made evident. Also, it helped us realize 
  there is an intrinsic angular momentum associated to each fundamental particle, known as spin. This was a crucial 
  discovery for the description of fundamental particles through the development of Quantum Field Theory 
  and their electromagnetic interactions by means of Quantum Electrodynamics (QED). \\
  
  The magnetic moment of a particle is defined to be the coupling strength between its electromagnetic current 
  and a magnetic field. As a result of this, one finds that the magnetic moment must be proportional to the angular momentum of the particle.
  One can compute the intrinsic magnetic 
  moment for fundamental particles that couple to the electromagnetic field 
  calculating their interaction with a classic electromagnetic field
  (as done by Dirac \cite{Dirac:1928ej}). This approach gives an intrinsic magnetic moment 
  \begin{equation}\label{Dirac}
  \mu = g\frac{q}{2m} s,
  \end{equation}
  where $q$ is the electric charge, $m$ is the mass of the particle, $s$ is the spin and $g=2$ is the gyromagnetic factor. 
%
 A precise measurement done by Isidor Isaac Rabi's group \cite{Nafe:1947zz} showed a deviation from the 
 value given by Dirac. This was explained by Julian Schwinger who computed the quantum correction to the interaction strength between the 
 electromagnetic current and the magnetic field, leading him to develop the necessary tools to renormalize QED in order to calculate
 the NLO correction \cite{Schwinger:1948iu}, $\delta\mu/\mu=\alpha/\pi+\mathcal{O}\left[(\alpha/\pi)^2\right]$, eliminating the incompatibility.
 The quantum corrections to $g=2$ 
  define the anomalous magnetic moment
  \begin{equation}
   a = \frac{g-2}{2}.
  \end{equation}
  Ever since, the anomalous magnetic moment of the electron, $a_e$, has been measured in evermore precise ways, 
  demanding more precise theoretical determinations of it.\\
  
  On the other hand, if one is interested in the search for Beyond Standard Model (BSM) effects in this observable 
  one has to take into account that dimension six operators will be proportional to the fermion mass divided by 
  heavy BSM scales. Since any observable depends on the squared modulus of the amplitude, such BSM effects will give 
  considerably larger contributions on heavier particles\footnote{The observable used for measuring $a_\mu$ is the $\mu$ decay width 
  $\Gamma(\mu\to\nu_\mu e \nu_e)$, where its polarization is known. This allows to measure the precession due to the interaction with
  the applied magnetic field.}. Being that the muon is $\sim 200$ times heavier than the electron, 
  BSM effects will yield a higher signal in $a_\mu$ than in $a_e$. These effects would be even higher 
  in the $\tau$ lepton, however $a_\tau$ is still compatible with zero\footnote{Although, there is an extraordinary 
  proposition for measuring $a_\tau$ by inserting a target inside the beampipe at the LHCb experiment, far from the main region of collisions. The produced $\tau$'s 
  would cross the pipe and enter a crystal where a sufficiently large electromagnetic field can be obtained, due to the potential 
  between crystalographic planes of a bent crystal, to give the precession of the lepton \cite{Joan Ruiz}. More details on the experimental 
  arrange are given in \cite{Bagli:2016mbf}.} \cite{PDG}. Hence, the study of intrinsic 
  magnetic moment of fundamental particles is still a very interesting subject nowadays.\\
  
  The current experimental value \cite{PDG} of $a_\mu = (11\hspace*{1ex}659\hspace*{1ex}209.1\pm 6.3)\cdot10^{-10}$, has been compared with very precise theoretical predictions. 
  These can be devided in three main parts, namely the QED part which contains contributions mainly from virtual leptons 
  and their electromagnetic interactions up to order $\left(\alpha/\pi\right)^5$ \cite{Kinoshita:2005sm}. This is the main contribution 
  to the total $a_\mu$. Nevertheless, its uncertainty, ${\Delta a_\mu}_{\rm QED}=0.008\cdot 10^{-10}$, is three orders of magnitude smaller than
  the experimental one. The second is the electroweak contribution which accounts for electroweak interactions excluding those which are pure 
  electromagnetic interactions. The computation of these up to two loops gives an uncertainty ${\Delta a_\mu}_{\rm EW}=0.10\cdot 10^{-10}$ \cite{PDG}, which 
  is still very small compared to the experimental one.\\
  
 \begin{figure*}
 \centering
 \includegraphics[scale=0.32]{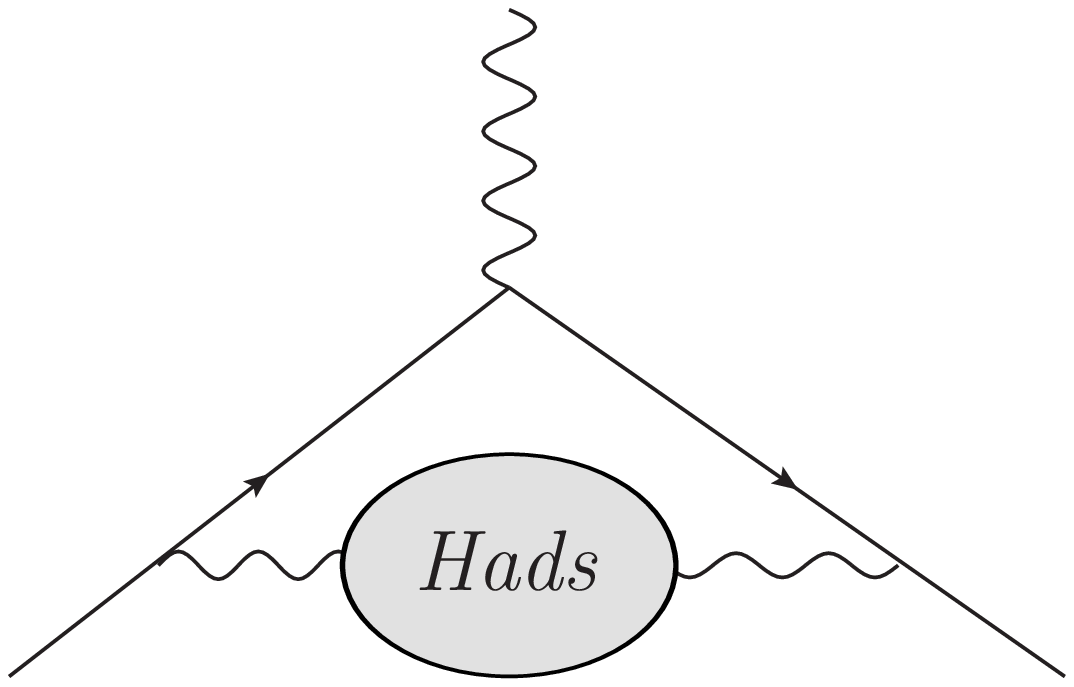}\hspace*{13ex}
 \includegraphics[scale=0.32]{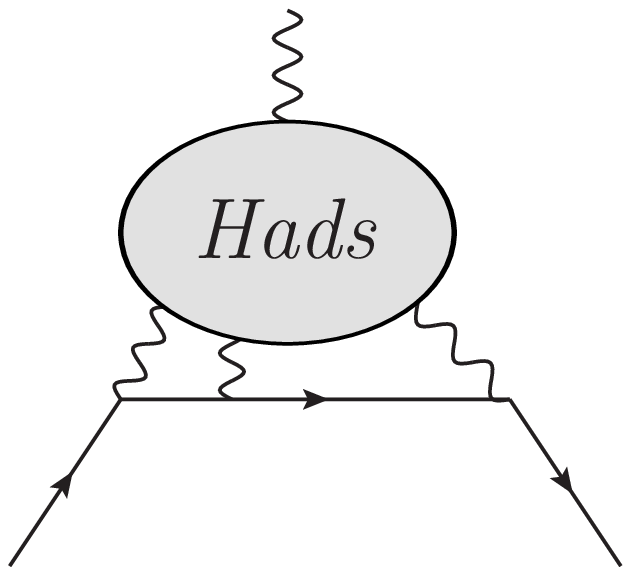}
 \caption{Hadronic contributions to $a_\mu$. The diagram on the left-hand-side represents all contributions from the
 hadronic vacuum to the self energy of the virtual photon, called Hadronic Vacuum Polarization (HVP). The diagram on the 
 right-hand-side represents all contributions from elastic scattering of two photons, called Hadronic Light-by-Light 
 scattering (HLbL).}
 \label{Hadronic_Part}       
 \end{figure*}  
  
  The remaining contributions are those containing quarks and strong interactions, these are 
  separated into two parts, the Hadronic Vacuum Polarization (HVP) and the Hadronic Light-by-Light scattering (HLbL), given in figure \ref{Hadronic_Part}. 
  The former can be extracted completely from experimental data on $R_{\rm had}=\sigma(e^+e^-\to {\rm hadrons})/\sigma(e^+e^-\to\mu^+\mu^-)$ and 
  contributes with an uncertainty ${\Delta a_\mu}_{\rm HVP}=3.4\cdot 10^{-10}$ \cite{PDG}; the latter cannot be fully obtained from experimental observables\footnote{
  See, however, the outstanding effort done in this direction from \cite{Bern,Bonn,Mainz}.} and needs to be obtained either numerically or on a model 
  dependent basis. However, this contributes with an error ${\Delta a_\mu}_{\rm HLbL}=2.6\cdot 10^{-10}$ \cite{PDG}. These uncertainties are 
  of the same order of magnitude as that given by the experiment. The most interesting fact is that the theoretical prediction is smaller  
  than the measured $a_\mu$, having an incompatibility\footnote{There is also an incompatibility between a recent measurement of $a_e$ 
  and the theoretical prediction, which uses a more precise determination of $\alpha$, of $\sim2.4\sigma$ \cite{Parker}. However, it is noteworthy that the 
  theoretical prediction is greater than the experimental one, contrary to the $a_\mu$ case.} of $\sim3.5\sigma$.  This has motivated new experiments aiming to increase the 
  precision in the determination of $a_\mu$, reducing the experimental error by, at least, a factor 4 in both, E34 at J-PARC \cite{Iinuma:2016zfu} and muon 
  g-2 at Fermilab \cite{Gohn:2017dsp}. Therefore, an effort must be done in the theoretical part to reduce the uncertainty by a similar factor. Since the
  HLbL part cannot be, nowadays, completely obtained from experiment, a deeper analysis of this part is necessary in order to reduce its uncertainty. \\

\begin{figure}[!ht]
   \centering
   \includegraphics[scale=0.45]{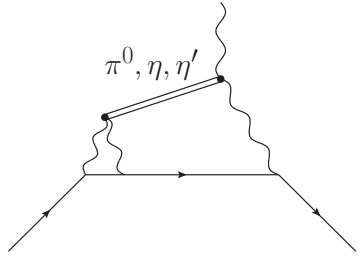}\caption{Main contribution to the Hadronic Light-by-Light piece of $a_\mu$.}\label{Pole diagram}
  \end{figure}

  This work is focused on the main contribution to the HLbL piece of the anomalous magnetic moment of the muon, $a_\mu^{HLbL}$, which is given by the 
  pseudoscalar exchange between pairs of photons, $a_\mu^{P,HLbL}$, \cite{Jegerlehner:2009ry} as shown in figure \ref{Pole diagram}. All that is needed to compute such contributions 
  to $a_\mu$ is the Transition Form Factor (TFF), $\mathcal{F}_{P\gamma^\star\gamma^\star}(q^2,p^2)$, of the pseudoscalar mesons coupling to two off-shell 
  photons with virtualities $q^2$ and $p^2$. 
  As has been shown in ref \cite{Knecht:2001qf}, $a_\mu^{P,HLbL}$ is almost fully determined by contributions at Euclidian squared photon momenta $-q^2,-p^2\lesssim 1$ GeV$^2$. 
  Therefore, it will be dominated mainly by the lowest-lying resonant part of the TFF and higher energies effects will give very small contributions. 
  To describe the pseudoscalar-TFF we rely on the extension of $\chi$PT \cite{ChPT} which incorporates the lightest resonances in a chiral 
  invariant way \cite{RCT}, namely Resonance Chiral Theory (R$\chi$T). Instead of using the complete basis of operators for resonances \cite{Kampf:2011ty}, 
  we will rely on the more simple 
  basis given in \cite{RuizFemenia:2003hm} to model vector meson interactions with pseudo-Goldstone bosons, since both are equivalent for describing vertices 
  involving only one pseudo-Goldstone
  \cite{Roig:2013baa}; nevertheless, we will use \cite{Kampf:2011ty} to account for pseudoscalar resonances effects. 
  The novelty in our approach is that we account for all the leading order terms that 
  break explicitly chiral symmetry, which enter as corrections in powers of the squared pseudo-Goldstone bosons masses, $m_P^2$. 

\section{Flavour {$U(3)$} breaking}
\label{sec-1}
 In this section, we will not show the full basis of operators (\cite{ChPT,RCT,Kampf:2011ty,RuizFemenia:2003hm,WZW,Bijnens:2001bb,Cirigliano:2006hb}), but
 will only show those which bring about the $U(3)$ breaking terms; the complete description is given in \cite{Guevara:2018rhj}.
 To consistently include all terms which break $U(3)$, an $\mathcal{O}(p^6)$ odd-intrinsic R$\chi$T Lagrangian with no resonances must be considered. 
 The contributions of $\mathcal{O}(p^4)$ will be given by the Wess-Zumino-Witten functional \cite{WZW}.
 The relevant non-resonant operators of $\mathcal{O}(p^6)$ are 
 \begin{eqnarray}\label{p6ChPT}
  {O}_7^W&=&i \epsilon_{\mu\nu\alpha\beta} \langle \chi_- f_+^{\mu\nu} f_+^{\alpha\beta}\rangle,\nonumber\\
  {O}_8^W&=&i \epsilon_{\mu\nu\alpha\beta} \langle \chi_-\rangle \, \langle  f_+^{\mu\nu} f_+^{\alpha\beta}\rangle,\nonumber\\
  {O}_{22}^W&=&i \epsilon_{\mu\nu\alpha\beta} \langle u^\mu \{ \nabla_\rho f_+^{\rho\nu}, f_+^{\alpha\beta}\}\rangle.
 \end{eqnarray}
 A correction to the vector resonance-photon coupling will be given by the interaction\footnote{This interaction term is 
 the only single-trace operator $\mathcal{O}(m_P^2)$ from those given in \cite{Cirigliano:2006hb}.}
 \begin{equation}
  \mathcal{L}_{VJ}=\frac{\lambda_V}{\sqrt{2}}\langle V_{\mu\nu}\{f^{\mu\nu}_+,\chi_+\}\rangle.
 \end{equation}
 There is also a correction to the mass of the vector resonances from V-V interactions\footnote{This term generates a mass spliting effect in the 
 nonet of resonances, inducing an explicit $U(3)$ breaking effect.}
 \begin{equation}
  \mathcal{L}_{VV}=-e_m^V\{V_{\mu\nu}V^{\mu\nu}\chi_+\}.
 \end{equation}
As a result, the masses of the vector resonances are given by
 \begin{equation}\label{VMasses}
  M_\rho^2=M_\omega^2 = M_V^2 - 4 e_m^V m_\pi^2,\hspace*{10ex} M_\phi^2=M_V^2 - 4 e_m^V \Delta_{2K\pi}^2,
 \end{equation}
 where $\Delta_{2K\pi}^2=2m_K^2-m_\pi^2$ and $M_V$ is the mass associated with the vector nonet in the chiral and large $N_C$ limits. 
 
 \section{Transition Form Factor}
 
 The Transition Form Factor (TFF) is defined through the $P\to\gamma^\star\gamma^\star$ decay amplitude, where the {\it dressing} of the 
 photons comes from the interaction with resonances and pseudo-Goldstones through their respective operators
 \begin{equation}
  \mathcal{M}_{P\to\gamma^\star\gamma^\star}=ie^2\varepsilon^{\mu\nu\alpha\beta}{q_1}_\mu{q_2}_\nu{\epsilon_1^*}_\alpha{\epsilon_2^*}_\beta
  \mathcal{F}_{P\gamma^\star\gamma^\star}(q_1^2,q_2^2),
 \end{equation}
 where $\epsilon_i = \epsilon(q_i)$ is the polarization of the photon with momentum $q_i$. Here, Bose symmetry implies 
 $\mathcal{F}_{P\gamma^\star\gamma^\star}(q_1^2,q_2^2)=\mathcal{F}_{P\gamma^\star\gamma^\star}(q_2^2,q_1^2)$. 
 One can impose relations among the parameters of the model by demanding that the TFF exhibits the short-distance behaviour expected from 
 QCD \cite{Brodsky:1973kr,Lepage:1980fj},
 \begin{equation}
  \lim_{q^2\to\infty}\mathcal{F}_{P\gamma^\star\gamma^\star}(q^2,q^2)=\mathcal{O}(q^{-2}){\rm\quad\quad and\quad\quad}
  \lim_{q^2\to\infty}\mathcal{F}_{P\gamma\gamma^\star}(0,q^2)=\mathcal{O}(q^{-2}).
 \end{equation}
 The full list of constraints obtained in this way for the parameters are shown in ref \cite{Guevara:2018rhj}. After applying the relations among 
 parameters, the simplified expression of the TFF for $\pi^0$ reads
  \begin{equation}\label{SimppiFF}
  \mathcal{F}_{\pi\gamma^\star\gamma^\star}(q_1^2,q_2^2)=\frac{32\pi^2m_\pi^2F_V^2d_{123}^\star -  N_C M_V^2M_\rho^2
   }{12 \pi ^2 F_\pi D_\rho(q_1^2)
   D_\rho(q_2^2)},
 \end{equation}
 where $F_\pi$ is the $\pi$ decay constant, $D_R(q^2)=M_R^2 - q^2$ is the denominator of the propagator of the vector-meson resonance $R$, 
 with the resonance masses $M_R$ given by (\ref{VMasses}),
 and $d_{123}^\star$ is a free parameter. Analogously, the simplified expression for the TFF of the $\eta$ is given by
  \begin{eqnarray}\label{SimpetaFF}
  \mathcal{F}_{\eta\gamma^\star\gamma^\star}(q_1^2,q_2^2)&=&\frac{1}{12\pi^2F D_\rho(q_1^2) D_\rho(q_2^2)D_\phi(q_1^2) D_\phi(q_2^2)}\times
  \\
   &&
   \hspace*{-2cm}\left\{-\frac{N_C M_V^2}{3     }\left[5C_q M_\rho^2D_\phi(q_1^2) D_\phi(q_2^2) - \sqrt{2}C_s M_\phi^2 D_\rho(q_1^2) D_\rho(q_2^2)\right]\right.
   \nonumber\\
   &&
   \hspace*{-2cm}+\frac{32\pi^2F_V^2d_{123}^\star m_\eta^2}{3}\left[(5C_q D_\phi(q_1^2) D_\phi(q_2^2)-\sqrt{2}C_s D_\rho(q_1^2) D_\rho(q_2^2)\right]
   \nonumber\\
   &&
   \hspace*{-2cm}\left.-\frac{256\pi^2F_V^2d_{2}^\star}{3}\left[(5C_q\Delta_{\eta\pi}^2 D_\phi(q_1^2) D_\phi(q_2^2)+\sqrt{2}C_s\Delta_{2K\pi\eta}^2 D_\rho(q_1^2) D_\rho(q_2^2)\right]\right\},\nonumber
 \end{eqnarray}
 where $d_2^\star$ is a free parameter, $\Delta_{\eta\pi}^2=m_\eta^2-m_\pi^2$, $\Delta_{2K\pi\eta}^2=2m_K^2-m_\pi^2-m_\eta^2$ and $C_{q/s}$ are the $\eta-\eta'$ 
 mixing parameters. The TFF for the $\eta'$ can be obtained from this by the substitutions $m_\eta\to m_{\eta'}$, $C_q\to C_q'$ and $C_s\to-C_s'$.\\
 
  As said previously, the evaluation of the contribution from the pseudo-Goldstone exchange is obtained by using the integral expressions given 
 in ref. \cite{Knecht:2001qf} substituting the TFF for each contribution. To get to these expressions, one has to assume that the form factor can 
 be expressed in the following way
 \begin{equation}\label{KnechtTFF}
  \mathcal{F}_{P\gamma^\star\gamma^\star}(q_1^2,q_2^2)= \frac{F}{3}\left[f(q^2) + \sum_{V_i}\frac{1}{M_{V_i}^2 - q_2^2}g_{V_i}(q_1^2)\right].
 \end{equation}
 After applying the short distance constraints, the function $f(q^2)$ vanishes for all the form factors, in accordance with previous determinations 
 of such function \cite{Knecht:2001qf,Roig:2014uja}.

 Since some of the parameters could not be constrained by imposing the correct high-energy behavior of the TFF, we fit them to experimental 
 determinations excluding the time-like ($q^2>0$) region of photon four-momenta, since radiative corrections might give large contributions 
 to the TFF in such region \cite{Husek:2017vmo}. We fitted simultaneously the parameters of our TFF of the $\pi^0$, $\eta$ and $\eta'$ 
 mesons to the decay widths of the three pseudo-Goldstones given by \cite{PDG}, also to the singly off-shell TFF from CELLO \cite{Behrend:1990sr} 
 and CELLO \cite{Gronberg:1997fj} for the three pseudo-Goldstones, 
 LEP for $\eta'$ \cite{Acciarri:1997yx}, BaBar for $\pi^0$ \cite{Aubert:2009mc}, BaBar for $\eta$ and $\eta'$ \cite{BABAR:2011ad} and Belle 
 for $\pi^0$ \cite{Uehara:2012ag}. All further details on the fit are given in ref \cite{Guevara:2018rhj}. \\
 
  \begin{figure}[!ht]
   \centering\includegraphics[scale=0.75]{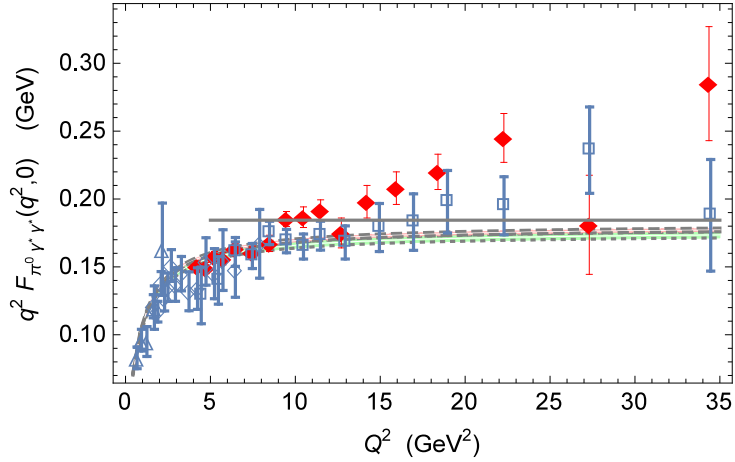}\caption{Fitted spacelike $\pi^0$-TFF. The red region shows the TFF using all data within 1-$\sigma$ 
     and the green region is the one excluding the BaBar data. The red diamonds are the BaBar data \cite{Aubert:2009mc}.}\label{fitted TFF}
  \end{figure}
 
 The fit including all data (fit1) gave a total $\chi^2/dof=150./101$. In comparison, the fit neglecting only the BaBar $\pi^0$ data from the whole set
 (fit 2), gave an improved value of $\chi^2/dof=69./84$, which we regarded as our best fit. The $\pi^0$-TFF prediction for both fits are shown in Fig. 
 \ref{fitted TFF}, where $Q^2=-q^2$ is the Euclidean squared momentum. 
 
 \section{Pseudo-Goldstone pole contribution to $a_\mu^{HLbL}$}
 
 \subsection{Meson exchange prediction with one vector resonance multiplet}
 
 The contribution from the pseudo-Goldstone pole to the HLbL piece of $a_\mu$, $a_\mu^{P,HLbL}$, is obtained using the integral representation 
 given in \cite{Knecht:2001qf}. The total pseudo-Goldstone contribution is estimated using a Monte Carlo run with $5\cdot10^{3}$ events which 
 randomly generates the eight fit parameters with a normal distribution according to their mean values, errors and correlations. The contributions
 from the three pseudo-Goldstones are integrated at the same time, accounting in this way for the correlation between the three contributions.
 Thus we obtain
 \begin{equation}
  a_\mu^{P,HLbL} = (8.47\pm0.16)\cdot10^{-10}.
 \end{equation}
 The prediction using the set of parameters from fit 1 is $a_{\mu}^{P,HLbL}=(8.58\pm0.16)\cdot10^{-10}$, which despite a higher central value 
 is completely compatible with the value we obtain using the parameters of fit 2. This is expected since one can see from Figure 
 \ref{fitted TFF} that the absolute value of the form factor is larger for this set of parameters at large $Q_i^2=-q_i^2$; however, since the integration 
 kernels are dominated by the region for $Q_i^2\lesssim1$ GeV$^2$ (as said above), the compatibility among both values is expected.\\
 
 We also study $a_\mu^{P,HLbL}$ by taking chiral and large $N_C$ limits of the TFF, keeping the physical masses in the integration kernels, 
 giving $(F/F_\pi)^2a_\mu^{P,LbL}=8.27\cdot10^{-10}$. This is obtained with the central values of the parameters of our best fit in these limits. Comparing 
 the latter with the central value of our contribution and taking $F\approx F_\pi$, we see that the chiral corrections account for a $\sim2.5$\%
 (up to corrections in $F/F_\pi$). This suggests that further chiral corrections (NNLO), suppressed by additional powers of $m_P^2$, must be negligible.\\
 
 \subsection{Further error analysis}
 
 The NLO effects in the $1/N_C$ expansion can be estimated by including the effects of the off-shell width in the $\rho$ meson propagator. The NLO 
 contributions to the latter are accounted mainly by the $\pi\pi$ and $\overline{K}K$ loops, the expression for such corrections reads \cite{GomezDumm:2000fz}
 \begin{equation}\label{NLO prop}
  M_\rho^2 - q^2 \quad\to\quad M_\rho^2 - q^2 + \frac{q^2M_\rho^2}{96\pi^2F_\pi^2}\left(A_\pi(q^2)+\frac{1}{2}A_K(q^2)\right),
 \end{equation}
 where the loop functions are given by
 \begin{equation}
  A_P(q^2)= \log\frac{m_P^2}{M_\rho^2} + 8\frac{m_P^2}{q^2} - \frac{5}{3} + \sigma_P^3(q^2)\log\left(\frac{\sigma_P(q^2) + 1}{\sigma_P(q^2) - 1}\right),
 \end{equation}
 being $\sigma_P(s)=\sqrt{1-\frac{4m_P^2}{s}}$. It is worth to notice that the loop functions are real for $q^2<4m_P^2$, so that the propagator is real 
 in the whole spacelike ($q^2<0$) region of photon momenta, where it is integrated. Since now the propagator of the $\rho$ meson is not a rational 
 function of $q^2$, it cannot be expressed as in eq. (\ref{KnechtTFF}). Therefore, in order to be able to express the TFF in such form we approximate 
 the form factor by imposing the condition obtained above that $f(q^2)$ vanishes and making the substitution (\ref{NLO prop}) in the rest of the expression
 in eq. (\ref{KnechtTFF}). This allows us to represent the TFF in such way that one can use the 
 integral representation in \cite{Knecht:2001qf} to obtain the $a_\mu^{P,HLbL}$ contribution. Thus, we obtain 
 ${a_\mu^{P,HLbL}|}_\text{LO+NLO}-{a_\mu^{P,HLbL}|}_\text{LO}
 = -0.09\cdot10^{-10}$.\\
 
 This is,
 nonetheless, just one of the possible NLO corrections in $1/N_C$ to the anomalous magnetic moment. One-loop modifications to the
 $\pi^0VV'$ vertex can be, e.g., equally important in the space-like domain and may lead to a positive con
 tribution to $a_\mu^{P,HLbL}$. Thus, we take the absolute value of this shift as a crude estimate of the $1/N_C$ effects:
 \begin{equation}
  \left(\Delta a_\mu^{P,HLbL}\right)_{1/N_C} = \pm 0.09\cdot10^{-10}.
 \end{equation}

 From the expressions (\ref{SimppiFF}) and (\ref{SimpetaFF}) it is evident that our TFF does not fulfill the exact short distance QCD limit expected 
 for $Q_1^2=Q_2^2=Q^2$ when $Q^2\to\infty$ \cite{Brodsky:1973kr,Lepage:1980fj}. Our form factors underestimate the real contribution since 
 they behave as $1/Q^4$ instead than $1/Q^2$ near this limit. One rough estimate can be given by computing the total contribution to $a_\mu^{P,HLbL}$ 
 with the form factors in the chiral limit with one and two vector resonance multiplets and comparing both results. The complete details of such 
 procedure are given in \cite{Guevara:2018rhj}. Thus, we obtain
 \begin{equation}
  \left(\Delta a_\mu^{P,HLbL}\right)_\text{asym} = ^{+0.5}_{-0.0}\cdot10^{-10}.
 \end{equation}
 \section{Conclusions}
 We have given a more accurate description of the TFF within the framework of R$\chi$T, including terms up to order $m_P^2$ for the first time in a chiral 
 invariant Lagrangian approach. This led to a more precise computation of the contribution from the $P$-pole to $a_\mu$. 
 By looking at the difference of our results with that using the TFF in the chiral limit ($0.20\cdot10^{-10}$) it seems that further chiral corrections 
 will be negligible. Considering all possible contributions to the error, we get
 \begin{equation}
  a_\mu^{P,HLbL}=(8.47\pm0.16_\text{stat}\pm0.09_{1/N_C}{^{+0.5}_{-0.0}}_\text{asym})\cdot10^{-10},
 \end{equation}
 where the first error (stat) comes from the fit, the second from possible $1/N_C$ corrections and the last due to the wrong asymptotic (asym) behavior of our TFF 
 estimated through the effect of heavier vector resonances.
 \acknowledgments
 This work was supported by CONACYT Projects No. FOINS-296-2016 (`Fronteras de la
Ciencia'), `Estancia Posdoctoral en el Extranjero' and 250628 (`Ciencia B\'asica'), and by the Spanish MINECO 
Project FPA2016-75654-C2-1-P.

\end{document}